\documentclass[prl,twocolumn,showpacs,linenumbers,preprintnumbers,amsmath,amssymb,superscriptaddress]{revtex4}

 \usepackage{graphicx}
 \usepackage{amsmath}
 \usepackage{amssymb}
 \usepackage{afterpage}
 \usepackage{graphics}
 \usepackage{float}
 \usepackage{rotating}
 \usepackage{xspace}
 \usepackage{dcolumn}%align table columns on decimal point
 \usepackage{bm} %bold math

\begin{document}


\title{Electron neutrino and antineutrino appearance in the full MINOS data sample}

\newcommand{\Berkeley}{Lawrence Berkeley National Laboratory, Berkeley, California, 94720 USA}
\newcommand{\Cambridge}{Cavendish Laboratory, University of Cambridge, Madingley Road, Cambridge CB3 0HE, United Kingdom}
\newcommand{\Cincinnati}{Department of Physics, University of Cincinnati, Cincinnati, Ohio 45221, USA}
\newcommand{\FNAL}{Fermi National Accelerator Laboratory, Batavia, Illinois 60510, USA}
\newcommand{\RAL}{Rutherford Appleton Laboratory, Science and Technologies Facilities Council, OX11 0QX, United Kingdom}
\newcommand{\UCL}{Department of Physics and Astronomy, University College London, Gower Street, London WC1E 6BT, United Kingdom}
\newcommand{\Caltech}{Lauritsen Laboratory, California Institute of Technology, Pasadena, California 91125, USA}
\newcommand{\Alabama}{Department of Physics and Astronomy, University of Alabama, Tuscaloosa, Alabama 35487, USA}
\newcommand{\ANL}{Argonne National Laboratory, Argonne, Illinois 60439, USA}
\newcommand{\Athens}{Department of Physics, University of Athens, GR-15771 Athens, Greece}
\newcommand{\NTUAthens}{Department of Physics, National Tech. University of Athens, GR-15780 Athens, Greece}
\newcommand{\Benedictine}{Physics Department, Benedictine University, Lisle, Illinois 60532, USA}
\newcommand{\BNL}{Brookhaven National Laboratory, Upton, New York 11973, USA}
\newcommand{\CdF}{APC -- Universit\'{e} Paris 7 Denis Diderot, 10, rue Alice Domon et L\'{e}onie Duquet, F-75205 Paris Cedex 13, France}
\newcommand{\Cleveland}{Cleveland Clinic, Cleveland, Ohio 44195, USA}
\newcommand{\Delhi}{Department of Physics \& Astrophysics, University of Delhi, Delhi 110007, India}
\newcommand{\GEHealth}{GE Healthcare, Florence South Carolina 29501, USA}
\newcommand{\Harvard}{Department of Physics, Harvard University, Cambridge, Massachusetts 02138, USA}
\newcommand{\HolyCross}{Holy Cross College, Notre Dame, Indiana 46556, USA}
\newcommand{\Houston}{Department of Physics, University of Houston, Houston, Texas 77204, USA}
\newcommand{\IIT}{Department of Physics, Illinois Institute of Technology, Chicago, Illinois 60616, USA}
\newcommand{\Iowa}{Department of Physics and Astronomy, Iowa State University, Ames, Iowa 50011 USA}
\newcommand{\Indiana}{Indiana University, Bloomington, Indiana 47405, USA}
\newcommand{\ITEP}{High Energy Experimental Physics Department, ITEP, B. Cheremushkinskaya, 25, 117218 Moscow, Russia}
\newcommand{\JMU}{Physics Department, James Madison University, Harrisonburg, Virginia 22807, USA}
\newcommand{\LASL}{Nuclear Nonproliferation Division, Threat Reduction Directorate, Los Alamos National Laboratory, Los Alamos, New Mexico 87545, USA}
\newcommand{\Lebedev}{Nuclear Physics Department, Lebedev Physical Institute, Leninsky Prospect 53, 119991 Moscow, Russia}
\newcommand{\LLL}{Lawrence Livermore National Laboratory, Livermore, California 94550, USA}
\newcommand{\LosAlamos}{Los Alamos National Laboratory, Los Alamos, New Mexico 87545, USA}
\newcommand{\Manchester}{School of Physics and Astronomy, University of Manchester, Oxford Road, Manchester M13 9PL, United Kingdom}
\newcommand{\MIT}{Lincoln Laboratory, Massachusetts Institute of Technology, Lexington, Massachusetts 02420, USA}
\newcommand{\Minnesota}{University of Minnesota, Minneapolis, Minnesota 55455, USA}
\newcommand{\Crookston}{Math, Science and Technology Department, University of Minnesota -- Crookston, Crookston, Minnesota 56716, USA}
\newcommand{\Duluth}{Department of Physics, University of Minnesota Duluth, Duluth, Minnesota 55812, USA}
\newcommand{\Ohio}{Center for Cosmology and Astro Particle Physics, Ohio State University, Columbus, Ohio 43210 USA}
\newcommand{\Otterbein}{Otterbein University, Westerville, Ohio 43081, USA}
\newcommand{\Oxford}{Subdepartment of Particle Physics, University of Oxford, Oxford OX1 3RH, United Kingdom}
\newcommand{\PennState}{Department of Physics, Pennsylvania State University, State College, Pennsylvania 16802, USA}
\newcommand{\PennU}{Department of Physics and Astronomy, University of Pennsylvania, Philadelphia, Pennsylvania 19104, USA}
\newcommand{\Pittsburgh}{Department of Physics and Astronomy, University of Pittsburgh, Pittsburgh, Pennsylvania 15260, USA}
\newcommand{\IHEP}{Institute for High Energy Physics, Protvino, Moscow Region RU-140284, Russia}
\newcommand{\Rochester}{Department of Physics and Astronomy, University of Rochester, New York 14627 USA}
\newcommand{\RoyalH}{Physics Department, Royal Holloway, University of London, Egham, Surrey, TW20 0EX, United Kingdom}
\newcommand{\Carolina}{Department of Physics and Astronomy, University of South Carolina, Columbia, South Carolina 29208, USA}
\newcommand{\SLAC}{Stanford Linear Accelerator Center, Stanford, California 94309, USA}
\newcommand{\Stanford}{Department of Physics, Stanford University, Stanford, California 94305, USA}
\newcommand{\StJohnFisher}{Physics Department, St. John Fisher College, Rochester, New York 14618 USA}
\newcommand{\Sussex}{Department of Physics and Astronomy, University of Sussex, Falmer, Brighton BN1 9QH, United Kingdom}
\newcommand{\TexasAM}{Physics Department, Texas A\&M University, College Station, Texas 77843, USA}
\newcommand{\Texas}{Department of Physics, University of Texas at Austin, 1 University Station C1600, Austin, Texas 78712, USA}
\newcommand{\TechX}{Tech-X Corporation, Boulder, Colorado 80303, USA}
\newcommand{\Tufts}{Physics Department, Tufts University, Medford, Massachusetts 02155, USA}
\newcommand{\UNICAMP}{Universidade Estadual de Campinas, IFGW-UNICAMP, CP 6165, 13083-970, Campinas, SP, Brazil}
\newcommand{\UFG}{Instituto de F\'{i}sica, Universidade Federal de Goi\'{a}s, CP 131, 74001-970, Goi\^{a}nia, GO, Brazil}
\newcommand{\USP}{Instituto de F\'{i}sica, Universidade de S\~{a}o Paulo,  CP 66318, 05315-970, S\~{a}o Paulo, SP, Brazil}
\newcommand{\Warsaw}{Department of Physics, University of Warsaw, Ho\.{z}a 69, PL-00-681 Warsaw, Poland}
\newcommand{\Washington}{Physics Department, Western Washington University, Bellingham, Washington 98225, USA}
\newcommand{\WandM}{Department of Physics, College of William \& Mary, Williamsburg, Virginia 23187, USA}
\newcommand{\Wisconsin}{Physics Department, University of Wisconsin, Madison, Wisconsin 53706, USA}
\newcommand{\deceased}{Deceased.}

\affiliation{\ANL}
\affiliation{\Athens}
%\affiliation{\Benedictine}
\affiliation{\BNL}
\affiliation{\Caltech}
\affiliation{\Cambridge}
\affiliation{\UNICAMP}
%\affiliation{\CdF}
\affiliation{\Cincinnati}
\affiliation{\FNAL}
\affiliation{\UFG}
\affiliation{\Harvard}
\affiliation{\HolyCross}
\affiliation{\Houston}
\affiliation{\IIT}
\affiliation{\Indiana}
\affiliation{\Iowa}
%\affiliation{\IHEP}
%\affiliation{\ITEP}
%\affiliation{\JMU}
%\affiliation{\Lebedev}
%\affiliation{\LLL}
\affiliation{\UCL}
\affiliation{\Manchester}
\affiliation{\Minnesota}
\affiliation{\Duluth}
\affiliation{\Otterbein}
\affiliation{\Oxford}
\affiliation{\Pittsburgh}
\affiliation{\RAL}
\affiliation{\USP}
\affiliation{\Carolina}
\affiliation{\Stanford}
\affiliation{\Sussex}
\affiliation{\TexasAM}
\affiliation{\Texas}
\affiliation{\Tufts}
\affiliation{\Warsaw}
%\affiliation{\Washington}
\affiliation{\WandM}
%\affiliation{\Wisconsin}

\author{P.~Adamson}
\affiliation{\FNAL}
%\affiliation{\UCL}
%\affiliation{\Sussex}

%\author{C.~Andreopoulos}
%\affiliation{\RAL}
%\affiliation{\Athens}

\author{I.~Anghel}
\affiliation{\ANL}
\affiliation{\Iowa}

%\author{K.~E.~Arms}
%\affiliation{\Minnesota}

%\author{R.~Armstrong}
%\affiliation{\Indiana}

%\author{D.~J.~Auty}
%\affiliation{\Sussex}

%\author{S.~Avvakumov}
%\affiliation{\Stanford}

%\author{D.~S.~Ayres}
%\affiliation{\ANL}

\author{C.~Backhouse}
\affiliation{\Oxford}

%\author{B.~Baller}
%\affiliation{\FNAL}

%\author{B.~Barish}
%\affiliation{\Caltech}

%\author{P.~D.~Barnes~Jr.}
%\affiliation{\LLL}

\author{G.~Barr}
\affiliation{\Oxford}

%\author{W.~L.~Barrett}
%\affiliation{\Washington}

%\author{E.~Beall}
%\altaffiliation[Now at\ ]{\Cleveland .}
%\affiliation{\ANL}
%\affiliation{\Minnesota}

%\author{B.~R.~Becker}
%\affiliation{\Minnesota}

%\author{A.~Belias}
%\affiliation{\RAL}

%\author{R.~H.~Bernstein}
%\affiliation{\FNAL}

%\author{M.~Betancourt}
%\affiliation{\Minnesota}

%\author{D.~Bhattacharya}
%\affiliation{\Pittsburgh}

%\author{M.~Bhattarai}
%\affiliation{\Texas}
%\affiliation{\Duluth}

\author{M.~Bishai}
\affiliation{\BNL}

\author{A.~Blake}
\affiliation{\Cambridge}

%\author{B.~Bock}
%\affiliation{\Duluth}

\author{G.~J.~Bock}
\affiliation{\FNAL}

%\author{D.~J.~Boehnlein}
%\affiliation{\FNAL}

\author{D.~Bogert}
\affiliation{\FNAL}

%\author{P.~M.~Border}
%\affiliation{\Minnesota}

%\author{C.~Bower}
%\affiliation{\Indiana}

%\author{E.~Buckley-Geer}
%\affiliation{\FNAL}

\author{S.~V.~Cao}
\affiliation{\Texas}

%\author{S.~Cavanaugh}
%\affiliation{\Harvard}

%\author{J.~D.~Chapman}
%\affiliation{\Cambridge}

\author{D.~Cherdack}
\affiliation{\Tufts}

% ...

\collaboration{The MINOS Collaboration}
\noaffiliation

\date{\today}

\date{\today}
\preprint{FERMILAB-PUB-13-023-E}

\begin{abstract}
We report on $\nu_e$ and $\bar{\nu}_e$ appearance in $\nu_\mu$ and $\bar{\nu}_\mu$ beams using the full MINOS data sample.  
The comparison of these $\nu_e$ and $\bar{\nu}_e$ appearance data at a 735 km baseline with $\theta_{13}$ measurements by reactor experiments probes $\delta$, 
the $\theta_{23}$ octant degeneracy, and the mass hierarchy.  This analysis is the first use of this technique and includes the first accelerator long-baseline search for $\bar{\nu}_\mu\rightarrow\bar{\nu}_e$.  
Our data disfavor 31\% (5\%) of the three-parameter space defined by $\delta$, the octant of the $\theta_{23}$, and the mass hierarchy at the 68\% (90\%) C.L.  
We measure a value of 2sin$^2(2\theta_{13})$sin$^2(\theta_{23})$ that is consistent with reactor experiments.
\end{abstract}

\maketitle

The neutrino oscillation phenomenon is successfully modeled by a theory of massive neutrino eigenstates that are different from the neutrino flavor eigenstates.  These sets of eigenstates are related by the PMNS matrix~\cite{ref:PNMS} which 
is commonly parameterized by three angles, $\theta_{ij}$, and a CP-violating phase, $\delta$.

The values of $\theta_{12}$ and $\theta_{23}$ have been measured~\cite{ref:SNO,ref:SK,ref:MINNOSMU} 
with indications that $\theta_{23}$ is not maximal~\cite{ref:Fogli,ref:Maltoni, ref:NeutrinoResults}.
The final angle, $\theta_{13}$, is now known 
to have a nonzero value from measurements by reactor experiments~\cite{ref:DC,ref:DB,ref:RENO}, the measurement by the T2K~\cite{ref:T2K} accelerator experiment, and from earlier MINOS results~\cite{ref:MINOSNUEOLD,ref:MINOSNUE}.

Despite these accomplishments, the value of $\delta$ is still unknown, as is the ordering of the neutrino masses, which is referred to as the neutrino mass 
hierarchy. Much of the attention in the neutrino community is now focused on resolving these unknowns. The mass hierarchy is not only a fundamental property of 
neutrinos but also has a direct impact on the ability of neutrinoless double beta decay searches to state definitively whether the neutrino is its own 
antiparticle~\cite{ref:doubleBeta}. Reactor experiments make a pure measurement of $\theta_{13}$, whereas the $\nu_\mu\rightarrow\nu_e$ and 
$\bar{\nu}_\mu\rightarrow\bar{\nu}_e$ appearance probabilities measured by accelerator experiments such as MINOS depend on the value of $\delta$ and 
sin$^2(\theta_{23})$.  In addition, the long-baseline of MINOS means that interactions between neutrinos and the matter of the Earth make the appearance probabilities dependent on the neutrino mass hierarchy~\cite{ref:msw,ref:3flav}.

We report the result from the search for $\nu_e$ ($\bar{\nu}_e$) 
appearance in a $\nu_\mu$ ($\bar{\nu}_\mu$) beam using the full MINOS 
data sample. This result uses an exposure of $10.6\times10^{20}$ 
protons-on-target taken with a $\nu$ beam and an exposure of 
$3.3\times10^{20}$ protons-on-target taken with a $\bar{\nu}$ beam.  The 
neutrino sample is 30\% larger than the sample used for the previous 
MINOS results on this topic~\cite{ref:MINOSNUE}.  This analysis 
represents the first long-baseline search for $\bar{\nu}_{\mu} 
\rightarrow \bar{\nu}_{e}$ appearance and places new constraints on 
$\theta_{13}$ and on a combination of $\delta$, 
$\theta_{23}$, and the neutrino mass hierarchy.

In the MINOS experiment~\cite{ref:minosnim}, neutrino oscillation is studied with the NuMI beamline~\cite{ref:beam} by measuring neutrino interactions in two detectors.  The Near Detector (ND), which has a fiducial mass of 29 tons, is at a distance of 1.04 km from the production target and is used to determine the composition of the beam before the neutrinos have oscillated. The Far Detector (FD), which has a fiducial mass of 3.8 kilotons, is at a distance of 735 km from the production target and is used to measure the change in the neutrino flavor composition of the beam.  In both the $\nu$ and $\bar{\nu}$ beam modes, the NuMI beam has an energy spectrum that is peaked at 3 GeV.  At the ND, the neutrino flavor composition of the neutrino interactions, as determined by a combination of simulation and measurement, is found to be 91.7\% $\nu_\mu$, 7.0\% $\bar{\nu}_\mu$, and 1.3\% $\nu_e$ and $\bar{\nu}_e$ for the $\nu$ beam mode and 58.1\% $\nu_\mu$, 39.9\% $\bar{\nu}_\mu$, and 2.0\% $\nu_e$ and $\bar{\nu}_e$ for the $\bar{\nu}$ beam mode.   

Both detectors are magnetized tracking calorimeters consisting of alternating planes of 2.54 cm thick steel and 1 cm thick scintillating plastic~\cite{ref:minosnim}.  The scintillator planes are segmented into 4.1 cm wide strips with wavelength-shifting fibers embedded in the strips to collect light for readout by multi-anode photomultiplier tubes.  

In the MINOS data sample, the flavor of a neutrino is determined only for charged-current (CC) interactions.  $\nu_{\mu}$-CC and $\bar{\nu}_{\mu}$-CC  interactions are identified by the presence of a long muon track that extends beyond a cluster of energy depositions that are consistent with hadronic activity at the interaction vertex.  Neutral-current (NC) interactions are identified by the energy depositions associated with hadronic activity.  $\nu_e$-CC and $\bar{\nu}_e$-CC  interactions produce an electromagnetic shower that typically leaves a compact cluster within 6 to 12 planes.  This analysis does not distinguish between $\nu_e$-CC and $\bar{\nu}_e$-CC interactions.
 
The sample of events classified as $\nu_e$-CC and $\bar{\nu}_e$-CC interactions contains 
a background of interactions with similar topology as required for  $\nu_{e}$-CC and $\bar{\nu}_{e}$-CC classification.  NC interactions with a significant 
electromagnetic component and $\nu_\mu$-CC or $\bar{\nu}_\mu$-CC interactions in which the muon track is not easily identified make up the majority of the 
background.  Smaller contributions to the background arise from $\nu_\tau$-CC and $\bar{\nu}_\tau$-CC interactions.  In addition to backgrounds that mimic $\nu_e$-CC and $\bar{\nu}_e$-CC event topologies, intrinsic $\nu_e$ and $\bar{\nu}_e$ components of the NuMI beam must be taken into account.

Candidate $\nu_e$-CC and $\bar{\nu}_e$-CC events are required to fall within a fiducial volume and 
to be coincident in time and direction with the NuMI beam.  We require the events to have 
shower-like topologies by rejecting events with tracks that are longer than 25 planes or extend 
more than 15 planes from a shower edge.  In addition, reconstructed events must have at least five 
consecutive planes with deposited energy above a threshold; this threshold is defined as half of the energy deposited by a minimum ionizing particle.  We require the events to have a reconstructed energy between 1 and 8 GeV where most of the $\nu_e$ and $\bar{\nu}_e$ appearance is expected.

\begin{figure}
\centering
\hfill \includegraphics[width=0.468\textwidth]{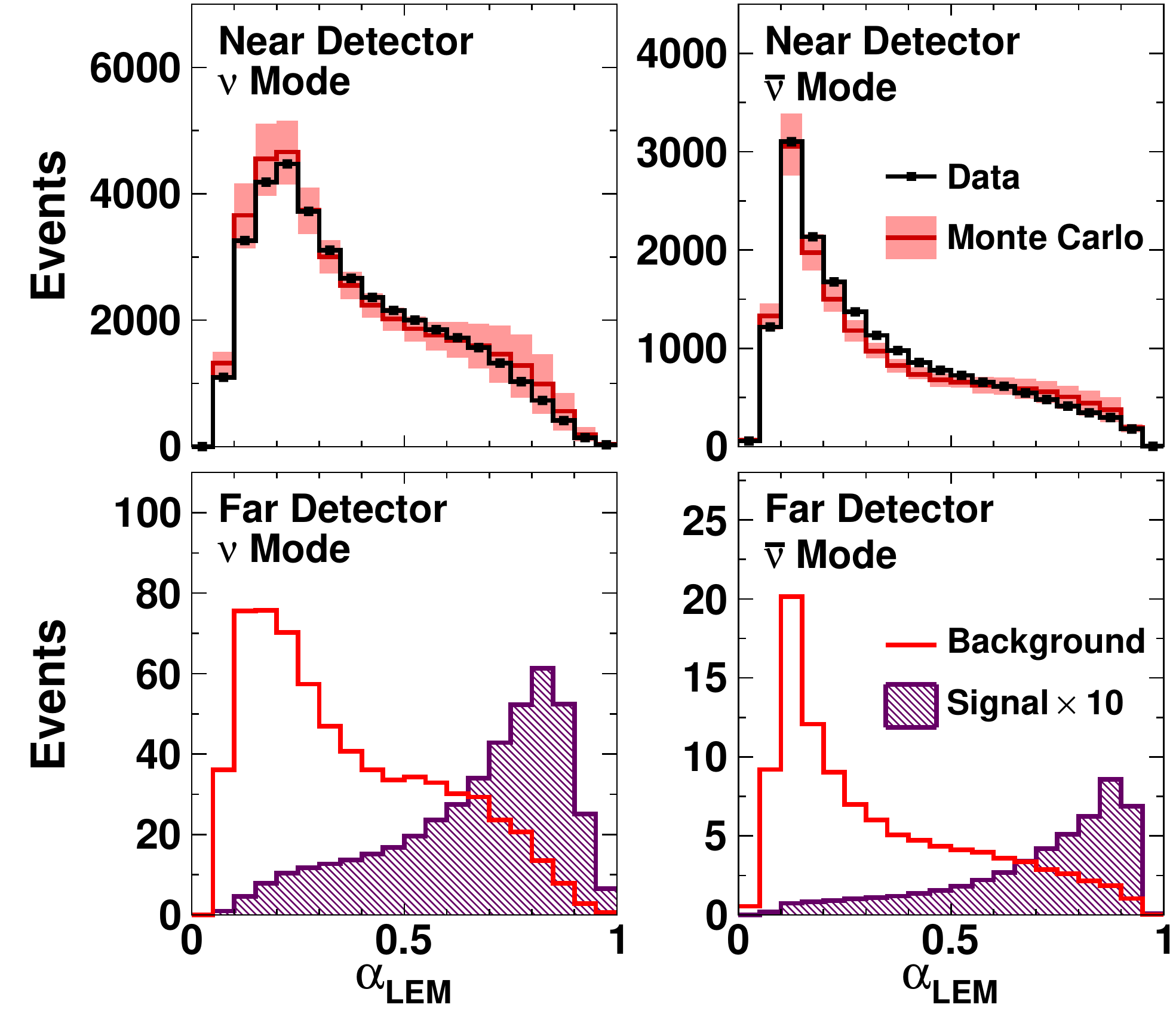}
\caption{Distributions of $\alpha_{LEM}$.  The plots in the left column correspond to the $\nu$ 
beam mode.  The plots in the right column correspond to the $\bar{\nu}$ beam mode.  The 
top row shows the distributions for ND selected events with a band about the simulation representing the 
systematic uncertainty.  The bottom row shows the distributions for the predicted FD background and 
signal multiplied by 10 with 2sin$^2(2\theta_{13})$sin$^2(\theta_{23})$ $=$ $0.1$, \mbox{$\delta$ 
$=$ 0}, and a normal mass hierarchy.}
\label{fig:lem}
\end{figure}

We further classify the events in this pre-selected sample of shower-like events by using a 
library-event-matching (LEM) algorithm~\cite{ref:pedro,ref:ruth}.  Within the LEM algorithm, the 
topology of energy depositions that characterize the event is compared to a library of simulated 
signal and background events.  Separate libraries are used for the events in the $\nu$ beam mode 
and $\bar{\nu}$ beam mode.  The 50 best-matching events in the library are collected and used to 
produce three variables.  These variables are the fraction of best-matching library events that are 
$\nu_{e}$-CC or $\bar{\nu}_{e}$-CC, the average inelasticity of the best-matching $\nu_{e}$-CC or 
$\bar{\nu}_{e}$-CC library events, and the average fraction of the energy depositions that overlap 
between the test event and the best-matching $\nu_{e}$-CC or $\bar{\nu}_{e}$-CC library events.  These three variables and the reconstructed neutrino energy of the test event are then used as an input into an artificial neural network.  The output value from the neural network is used to discriminate between signal and background events. This discriminant variable is referred to as $\alpha_{LEM}$ and is shown in Fig.~\ref{fig:lem}.  Signal events have a value near one, while background events cluster near zero.  The maximum sensitivity to $\nu_e$ and $\bar{\nu}_e$ appearance is obtained by analyzing events with \mbox{$\alpha_{LEM}$ $>$ 0.6}.

Following the selection of $\nu_{e}$-CC and $\bar{\nu}_{e}$-CC candidate events, the ND data are used to study the rate of background from NC, $\nu_{\mu}$-CC 
and $\bar{\nu}_{\mu}$-CC and intrinsic beam $\nu_{e}$-CC and $\bar{\nu}_{e}$-CC interactions. The NuMI beam can be tuned to produce different energy spectra. Among these different beam configurations, the relative contributions of the various backgrounds change in a well-understood way. By measuring the total of the three backgrounds in three different beam configurations, the relative amounts of the individual backgrounds can be deduced~\cite{ref:joao}.

We use the measurement of the ND backgrounds to derive the FD background predictions for the data samples in the $\nu$ beam mode and in the $\bar{\nu}$ beam mode.  For each sample, we divide simulated FD events into bins of energy and $\alpha_{LEM}$ and correct each FD background component, bin-by-bin, by multiplying it by the measured ND ratio of data to simulated events for that background.
  Since the ND data sample does not contain $\nu_{\tau}$-CC and $\bar{\nu}_{\tau}$-CC events from oscillation, we estimate the FD contribution from this small background component through simulation and a correction based on the observed ND $\nu_{\mu}$-CC and $\bar{\nu}_{\mu}$-CC spectra.

The sources of systematic uncertainty that affect the background prediction are given in 
Table~\ref{tab:systematics}. The effect of each source of uncertainty is evaluated by producing simulated ND and FD event samples that are modified according to the estimated size of each systematic effect. These modified samples are used to produce an altered FD background prediction for the systematic effect in question. We take the resulting difference between the nominal and modified prediction as the systematic uncertainty on the background prediction.
The systematic effect that results in the largest reduction in sensitivity is a 2.0\% uncertainty on the relative energy scale between the ND and FD.

\begin{table}
\begin{tabular}{l c c}
\hline 
\hline 
Systematic Effect & Uncertainty & Uncertainty \\
 & $\nu$ mode & $\bar{\nu}$ mode \\
\hline 
Energy Scale &2.7\% & 3.0\%\\
Normalization  & 1.9\% & 1.9\%\\
$\nu_\tau$  cross-section & 1.7\% & 2.0\%\\
All Others & 0.8\% & 2.5\%\\
\hline
Total Systematic & 3.8\%  & 4.8\% \\
\hline
Total Statistical & 8.8\%  & 23.9\% \\
\hline	
\hline
\end{tabular}
\caption{Systematic uncertainty on the FD background prediction for events with a value of 
\mbox{$\alpha_{LEM}$ $>$ 0.6}.  Effects listed under ``All Others'' include the neutrino flux, 
cross 
sections, detector modeling, and background decomposition.}
\label{tab:systematics}
\end{table}

With the absence of a $\nu_e$-CC and $\bar{\nu}_e$-CC signal in the ND, the signal selection efficiency cannot be extrapolated from the ND events in the same way as the background estimate. Therefore, to evaluate the signal efficiency, we select a sample of well identified $\nu_{\mu}$-CC events~\cite{ref:rustem,ref:cc2010}, remove the energy depositions that are associated with the muon track~\cite{ref:ANNA}, and insert the simulated energy depositions of an electron with an identical 
three-momentum~\cite{ref:JOSH}.  This method effectively turns a well identified sample of $\nu_\mu$-CC and 
$\bar{\nu}_\mu$-CC data events into a sample of $\nu_e$-CC and $\bar{\nu}_e$-CC data events.  For 
the $\nu$ beam mode ($\bar{\nu}$ beam mode) data sample, we find the expected number of FD signal events with $\alpha_{LEM}$ 
$>$ 0.6 and the associated systematic uncertainty to be $33.7 \pm 1.9$ ($3.9 \pm 0.2$) assuming 
sin$^2(2\theta_{13})$ = 0.1, \mbox{$\delta$ = 0}, $\theta_{23}$ = $\pi/4$, and a normal mass 
hierarchy.  This corresponds to an identification efficiency of 
(57.4 $\pm$ 2.8)\% for the $\nu$ beam mode and of (63.3 $\pm$  3.1)\% for the $\bar{\nu}$ beam mode.  The systematic uncertainties are evaluated 
in a way that is similar to the evaluation of the background systematics by using simulated samples that have been altered by a systematic effect.

Events with \mbox{$\alpha_{LEM}$ $<$ 0.5} are insensitive to $\nu_e$ and $\bar{\nu}_e$ appearance.  
These 
events are therefore used in a separate study to validate the analysis procedure. ND events with 
\mbox{$\alpha_{LEM}$ $<$ 0.5} are used to predict FD event yields, which are found to agree with 
the 
FD 
data to within 0.3 (0.6) standard deviations of the statistical uncertainty for the data sample in the $\nu$ ($\bar{\nu}$) beam mode.  

\begin{figure}
\centering
\hfill \includegraphics[width=0.44\textwidth]{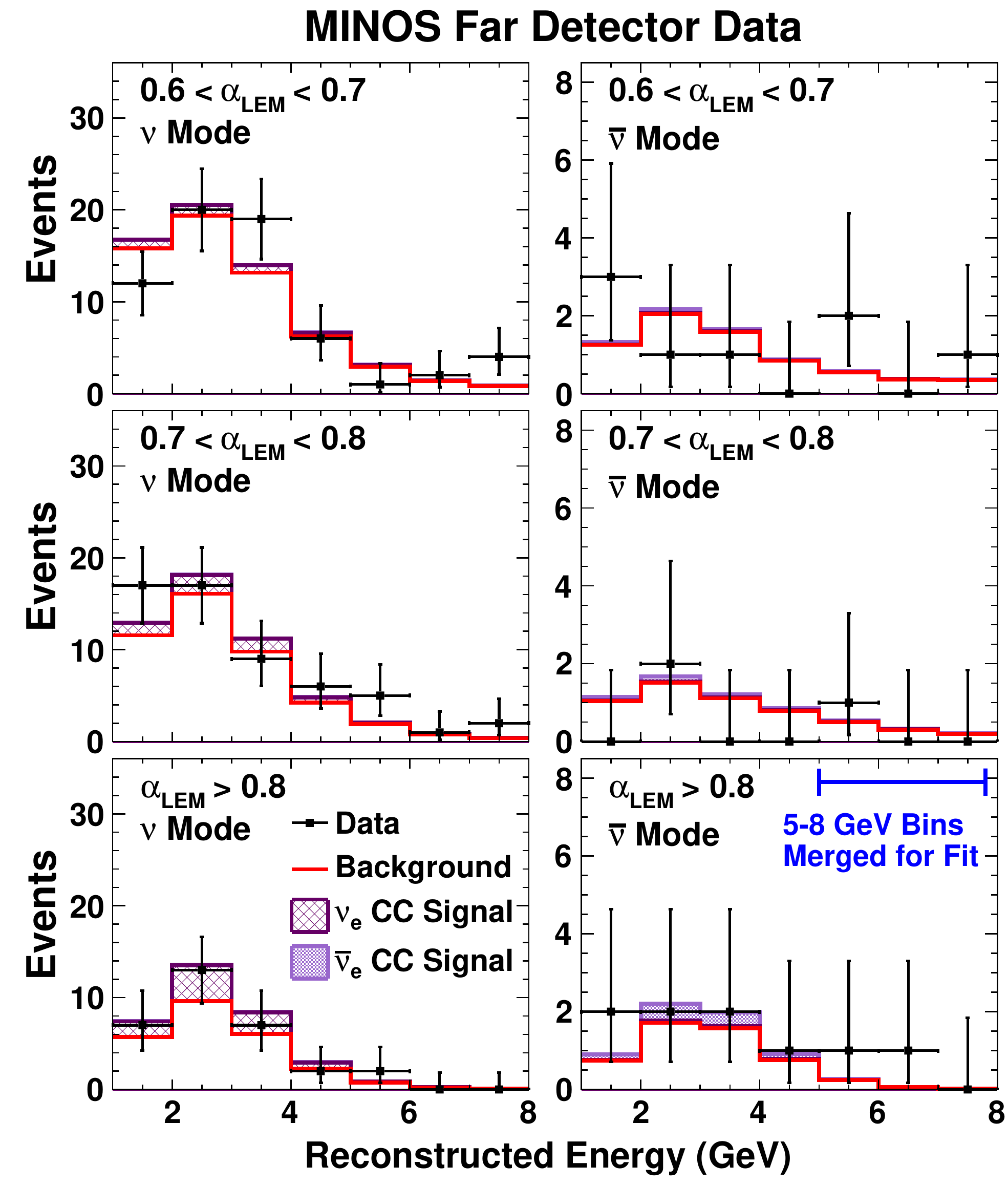}
\caption{The reconstructed energy distributions for three $\alpha_{LEM}$ ranges.  The events with 
energy greater than 5 GeV are combined into a single bin for the fits.  The vertical bars through 
the data points denote statistical uncertainties.  The signal predictions assume 
sin$^2$(2$\theta_{13}$) = 0.051, $\Delta m^{2}_{32}$ $>$ 0, $\delta$ = 0, and \mbox{$\theta_{23}$ 
= $\pi/4$}.  
The plots in the left column correspond to data collected in the $\nu$ beam mode.  The plots in the 
right column correspond to data collected in the $\bar{\nu}$ beam mode.}
\label{fig:fddata}
\end{figure}

\begin{figure}
\centering
\hfill \includegraphics[width=0.48\textwidth, trim = 10 0 0 0, 
clip]{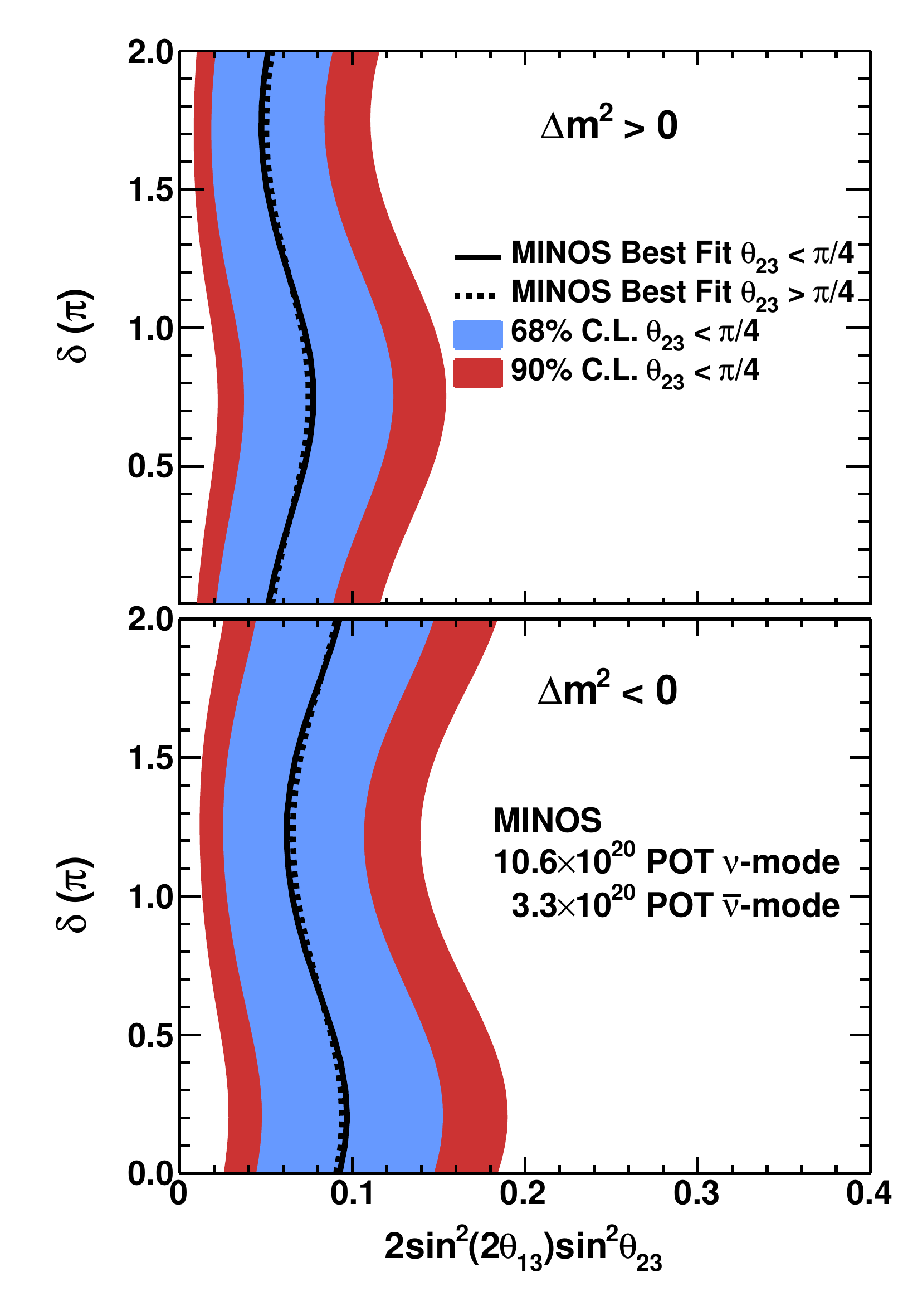}
\caption{The 68\% and 90\% confidence intervals of allowed values for 2sin$^2(2\theta_{13})$sin$^2(\theta_{23})$ as a function of $\delta$ for the two mass hierarchies.}
\label{fig:contour}
\end{figure}

Events with \mbox{$\alpha_{LEM}$ $>$ 0.6} are selected for further analysis in 
the 
$\nu$ beam mode and in the $\bar{\nu}$ beam mode. The expected and observed event counts in 
these samples are shown in Table~\ref{tab:events}. The observed FD reconstructed energy spectra, in bins of $\alpha_{LEM}$, are shown for the candidate events 
in Fig.~\ref{fig:fddata}.  Assuming a three-flavor neutrino oscillation probability that includes matter effects~\cite{ref:3flav}, we simultaneously fit the 
data from the $\nu$ beam mode and $\bar{\nu}$ beam mode samples for the value of 2sin$^2(2\theta_{13})$sin$^2(\theta_{23})$ while the value of the mass 
hierarchy and $\delta$ are held fixed.  The fit is performed using the 15 bins formed by three bins of $\alpha_{LEM}$ and five bins of energy.  This procedure is performed for all values of $\delta$ and both mass hierarchies, and the resulting confidence intervals, calculated using the Feldman-Cousins technique~\cite{ref:FC}, are shown in Fig.~\ref{fig:contour}.  The values of the oscillation parameters used in the fit are taken from previous measurements~\cite{ref:MINNOSMU,ref:SNO} and are set to 
\mbox{sin$^{2}(2\theta_{23})$ = 0.957 $^{+0.035}_{-0.036}$}, 
\mbox{$|\Delta m^{2}_{32}|$ = $(2.39$ $^{+0.09}_{-0.10})\times10^{-3}$ eV$^2$}, 
\mbox{$\theta_{12}$ = 0.60 $\pm$ 0.02},
 and \mbox{$\Delta m^{2}_{21}$ = $(7.59$ $^{+0.19}_{-0.21})\times10^{-5}$ eV$^2$}.  The full set of statistical and systematic uncertainties on the prediction are taken into account when constructing the contours.

Assuming a normal mass hierarchy, \mbox{$\delta$ = 0}, and 
\mbox{$\theta_{23}<\pi/4$}, we find that the data allow for values of 
\mbox{0.01 $<$ 2sin$^2(2\theta_{13})$sin$^2(\theta_{23})$ $<$ 0.12} at 90\% C.L. with the best-fit value of 
\mbox{2sin$^2(2\theta_{13})$sin$^2(\theta_{23})$ $=$ $0.051^{+0.038}_{-0.030}$}.  Assuming an inverted mass hierarchy, \mbox{$\delta$ = 0}, and \mbox{$\theta_{23}<\pi/4$}, we find that the data allow for values of 
\mbox{0.03 $<$ 2sin$^2(2\theta_{13})$sin$^2(\theta_{23})$ $<$ 0.18} at 90\% C.L. with the best-fit value of 
\mbox{2sin$^2(2\theta_{13})$sin$^2(\theta_{23})$ $=$ $0.093^{+0.054}_{-0.049}$}.  The best-fit 
values show very weak dependence on the choice of octant for $\theta_{23}$.

\begin{table}
\begin{tabular}{l c r}
\hline 
\hline 
Event Type & $\nu$ beam & $\bar{\nu}$ beam \\
& mode & mode \\
\hline 
NC 				   		 					&89.4 & 13.9 \\
$\nu_{\mu}$-CC and $\bar{\nu}_{\mu}$-CC  		 &21.6 & 1.0 \\
Intrinsic $\nu_{e}$-CC and $\bar{\nu}_{e}$-CC      &11.9 & 1.8 \\
$\nu_{\tau}$-CC and $\bar{\nu}_{\tau}$-CC  	 &4.8 & 0.8 \\
$\nu_\mu \rightarrow \nu_{e}$-CC 	         &33.0 & 0.7 \\
$\bar{\nu}_\mu \rightarrow \bar{\nu}_{e}$-CC     &0.7 & 3.2 \\
\hline
Total & 161.4  & 21.4 \\
\hline
Data & 152  & 20 \\
\hline
\hline
\end{tabular}
\caption{Expected FD event yields for events with a value of \mbox{$\alpha_{LEM}$ $>$ 0.6}, 
assuming 
sin$^2(2\theta_{13})$ = 0.1, $\delta$ = 0, $\theta_{23}$ = $\pi/4$, and a normal mass hierarchy.}
\label{tab:events}
\end{table}

\begin{figure}
\centering
\hfill \includegraphics[width=0.44\textwidth, trim = 38 0 80 60, clip]{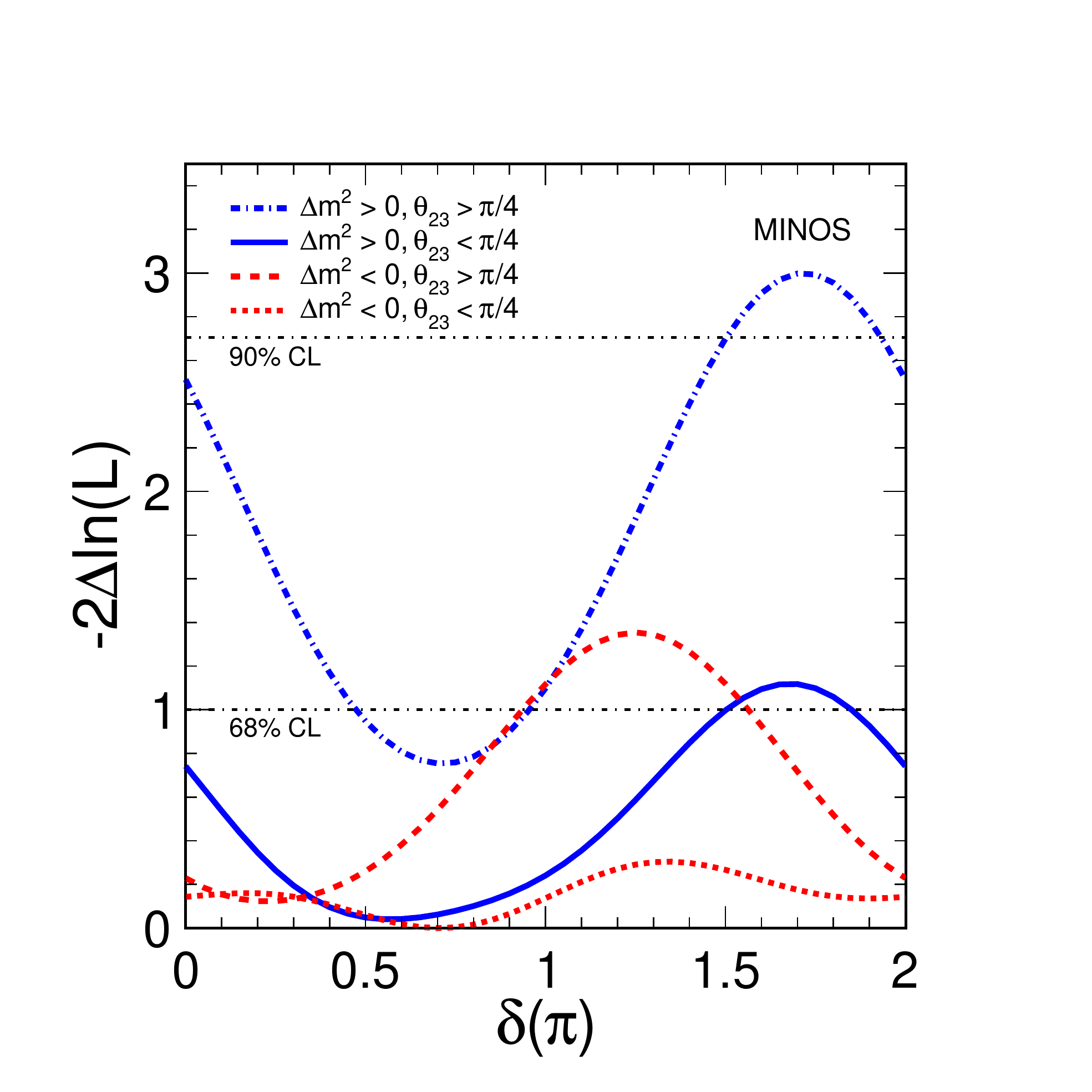}
\caption{The resulting values of the likelihood $L$, shown here as $-2\Delta$ln$L$, from a fit of $\delta$ to our data using constraints from reactor 
experiments~\cite{ref:DC,ref:DB,ref:RENO}, assuming various values of the mass hierarchy and the sign of $\theta_{23}-\pi/4$.  The difference is taken with 
respect to the best-fitting solution.  Values above the horizontal dashed lines are disfavored at either 68\% or 90\% C.L.}
\label{fig:delta}
\end{figure}

We are further able to place constraints on the value of $\delta$, the octant of $\theta_{23}$, and the neutrino mass hierarchy by incorporating the current 
knowledge of \mbox{$\sin^2(2\theta_{13})$ $=$ $0.098\pm0.013$} that we calculate from recent reactor data~\cite{ref:DC,ref:DB,ref:RENO}.  Fig.~\ref{fig:delta} 
shows the likelihood for our data as a function of 
$\delta$ for the four possible combinations of mass hierarchy and the octant of $\theta_{23}$.  The full set of statistical and systematic uncertainties on the prediction are taken into account when calculating the likelihood, 
as are the uncertainties on the oscillation parameters.  This analysis represents the first result by a long-baseline experiment to use a combination of $\nu_\mu\rightarrow\nu_e$ and $\bar{\nu}_\mu\rightarrow\bar{\nu}_e$ 
appearance data, with external reactor constraints on $\theta_{13}$, to probe $\delta$, the $\theta_{23}$ octant degeneracy, and the mass hierarchy.  Assuming 
$\theta_{23} > \pi/4$ ($\theta_{23} < \pi/4$), the data prefer 
an inverted hierarchy at 0.63 (0.04) units of $-2\Delta$ln$L$.  Furthermore, as is indicated by the horizontal lines in Fig.~\ref{fig:delta}, our data disfavor 31\% (5\%) of the three-parameter space defined by $\delta$, the 
octant of the $\theta_{23}$, and the mass hierarchy at the 68\% (90\%) C.L.  This analysis demonstrates the potential discriminating power that can be achieved with the combination of reactor and $\nu_\mu\rightarrow\nu_e$ and 
$\bar{\nu}_\mu\rightarrow\bar{\nu}_e$ appearance data.

 In conclusion, we have presented the results of $\nu_e$ and $\bar{\nu}_e$ appearance in $\nu_\mu$ and $\bar{\nu}_\mu$ beams from the full MINOS 
data sample.  
 We have used these data to place new constraints on the mixing angle $\theta_{13}$ and have demonstrated how such data will be used in the future to break the 
degeneracy in the appearance probability created by the ambiguity in the octant of $\theta_{23}$, the neutrino mass hierarchy, and the value of the CP-violating phase $\delta$.

This work was supported by the US DOE; the UK STFC; the US NSF; the State and University of 
Minnesota; the University of Athens, Greece; and Brazil's FAPESP, CNPq, and CAPES.  We are 
grateful to the Minnesota DNR, the crew of the Soudan Underground Laboratory, and
the personnel of Fermilab for their contributions to this effort.  We thank Texas Advanced Computing Center at The University of Texas at Austin for the provision of computing resources.


\begin{thebibliography}{99}
\bibitem{ref:PNMS}{B. Pontecorvo, JETP {\bf 34}, 172 (1958); V. N. Gribov and B. Pontecorvo, Phys. Lett. B {\bf 28}, 493 (1969); Z. Maki, M.  Nakagawa, and S. Sakata, Prog. Theor. Phys. {\bf 28}, 870 (1962).}

\bibitem{ref:SNO} B. Aharmim {\it et al.}  (SNO), Phys. Rev. Lett. {\bf 101}, 111301 (2008).
\bibitem{ref:SK} Y. Ashie {\it et al.}  (Super-Kamiokande), Phys. Rev. D {\bf 71}, 112005 (2005).
\bibitem{ref:MINNOSMU} R. Nichol, in {\it Proceedings of the XXV International Conference on Neutrino Physics and Astrophysics, Kyoto, Japan, June 2012} (to be published).

\bibitem{ref:Fogli} G.~L.~Fogli, E.~Lisi, A.~Marrone, D.~Montanino, A.~Palazzo, and A.~M.~Rotunno, Phys. Rev. D {\bf 86}, 013012 (2012).


\bibitem{ref:Maltoni} M.~C.~Gonzalez-Garcia, Michele Maltoni, Jordi Salvado, and Thomas Schwetz, JHEP {\bf 12}, 123 (2012).
\bibitem{ref:NeutrinoResults} Y. Itow, in {\it Proceedings of the XXV International Conference on Neutrino Physics and Astrophysics, Kyoto, Japan, June 2012} (to be published).



\bibitem{ref:DC} Y. Abe {\it et al.}   (Double Chooz), Phys. Rev. Lett. {\bf 108}, 131801 (2012).
\bibitem{ref:DB} F. P. An {\it et al.}   (Daya Bay), Phys. Rev. Lett. {\bf 108}, 171803 (2012).
\bibitem{ref:RENO} J. K. Ahn {\it et al.}   (RENO), Phys. Rev. Lett. {\bf 108}, 191802 (2012).

\bibitem{ref:T2K} K. Abe {\it et al.}   (T2K), Phys. Rev. Lett. {\bf 107}, 041801 (2011).

\bibitem{ref:MINOSNUEOLD} P. Adamson {\it et al.}   (MINOS), Phys. Rev. Lett. {\bf 103}, 261802 
(2009); Phys. Rev. D {\bf 82}, 051102R (2010).

\bibitem{ref:MINOSNUE} P. Adamson {\it et al.}   (MINOS), Phys. Rev. Lett. {\bf 107}, 181802 (2011).


\bibitem{ref:doubleBeta}{J. Beringer {\it et al.}    (PDG), Phys. Rev. D {\bf 86}, 010001 (2012).}

\bibitem{ref:msw}{L. Wolfenstein, Phys. Rev. D {\bf 17}, 2369 (1978); S. P. Mikheyev and A. Yu. Smirnov, Sov. J. Nucl. Phys. {\bf 42},
913 (1985).}	

\bibitem{ref:3flav}{E. K. Akhmedov {\it et al.}, J. High En. Phys. {\bf 04}, 078 (2004).}
\bibitem{ref:minosnim}{D. G. Michael {\it et al.}  (MINOS), Nucl. Inst. \& Meth. A {\bf 596}, 190 
(2008). 
}
\bibitem{ref:beam} K. Anderson {\it et al.}, FERMILAB-DESIGN-1998-01 (1998).

\bibitem{ref:pedro}{J. P. Ochoa, Ph.D. Thesis, California Institute of Technology, FERMILAB-THESIS-2009-44 (2009). }
\bibitem{ref:ruth}{R. Toner, Ph.D. Thesis, University of Cambridge, FERMILAB-THESIS-2011-53  (2011). }

\bibitem{ref:joao}{J. A. B. Coelho, Ph.D. Thesis, Universidade Estadual de Campinas, FERMILAB-THESIS-2012-23  (2012).}

\bibitem{ref:rustem}{R. Ospanov, Ph.D. Thesis, University of Texas at Austin, FERMILAB-THESIS-2008-04 (2008).}
\bibitem{ref:cc2010} P. Adamson {\it et al.}   (MINOS), Phys. Rev. Lett. {\bf 106}, 181801 (2011).

\bibitem{ref:ANNA}{A. Holin, Ph.D. Thesis, University College London, FERMILAB-THESIS-2010-41 (2010).}
\bibitem{ref:JOSH}{J. Boehm, Ph.D. Thesis, Harvard University, FERMILAB-THESIS-2009-17  (2009).}

\bibitem{ref:FC} {G. J. Feldman and R. D. Cousins, Phys. Rev. D {\bf 57}, 3873 (1998).}


\end{thebibliography}
\end{document}